\documentclass{INTERSPEECH2023}
\interspeechcameraready
\usepackage{cite}

\title{eCat: An End-to-End Model for Multi-Speaker TTS \& Many-to-Many Fine-Grained Prosody Transfer}
\name{
    Ammar Abbas$^*$, 
    Sri Karlapati$^*$\thanks{$^*$Equal contribution. Order determined by tossing a coin.}, 
    Bastian Schnell,
    Penny Karanasou,
    Marcel Granero Moya, \\
    Amith Nagaraj,
    Ayman Boustati,
    Nicole Peinelt,
    Alexis Moinet,
    Thomas Drugman
}   
\address{
    Alexa AI, Amazon
}
\email{syeabbs@amazon.co.uk, srikarla@amazon.co.uk}

\begin{document}

    \maketitle
 
    \begin{abstract}
        We present eCat, a novel end-to-end multispeaker model capable of: a) generating long-context speech with expressive and contextually appropriate prosody, and b) performing fine-grained prosody transfer between any pair of seen speakers. eCat is trained using a two-stage training approach. In Stage~I, the model learns speaker-independent word-level prosody representations in an end-to-end fashion from speech. In Stage~II, we learn to predict the prosody representations using the contextual information available in text. We compare eCat to CopyCat2, a model capable of both fine-grained prosody transfer~(FPT) and multi-speaker TTS. We show that eCat statistically significantly reduces the gap in naturalness between CopyCat2 and human recordings by an average of $46.7\%$ across $2$ languages, $3$ locales, and $7$ speakers, along with better target-speaker similarity in FPT. We also compare eCat to VITS, and show a statistically significant preference.
    \end{abstract}
    
    \vspace{-1mm}
    \section{Introduction}
        Previously, Neural Text-to-Speech (NTTS) methods depended on training a separate component to generate mel-spectrograms from phonemized text~\cite{tacotron2, li2018close, skerry2018towards, fastspeech2} and another component (vocoder) to generate speech waveforms from mel-spectrograms~\cite{UniversalWavenet, lorenzo2018robust, kumar2019melgan, kong2020hifigan, lee2023bigvgan}. The mel-spectrograms served as an intermediary between the two components. This approach suffered from compounding errors, as the vocoder is exposed to mel-spectrograms obtained from speech waveforms during training but predicted mel-spectrograms during inference. To assuage this, the vocoder was also sometimes fine-tuned on synthesised mel-spectrograms. Recently, NTTS methods have begun to use a single end-to-end model to generate speech waveforms directly from phonemized input text~\cite{vits-kim21f,tan22naturalspeech,donahue22eats}, resulting in an improvement in the naturalness of synthesised speech. 
        
        On a parallel note, there has also been work in multi-sentence TTS or long-context TTS~\cite{makarov22blt,xue22paratts,xin22audiobooktts,pan21chaptts}. The aim of such methods is to generate speech with consistent inter-sentence prosody by exploiting the dependency between consecutive sentences in long-context text. Makarov \emph{et~al.}~\cite{makarov22blt} concatenate consecutive sentences into a single training sentence and train the TTS system to predict the mel-spectrogram for the concatenated sentences. Detai \emph{et~al.}~\cite{xin22audiobooktts} also use acoustic and textual context from surrounding sentences to synthesise speech with contextually appropriate prosody. Liumeng \emph{et~al.}~\cite{xue22paratts} show that learning inter-sentence information in paragraphs with multi-head attention mechanism improves naturalness of TTS over sentence-based models. Overall, these methods affirm that using additional context around the target sentence improves the contextual relevance of prosody across sentences.
        
        In the realm of prosody representation learning, there has been significant work in learning speaker independent prosody representations for many-to-many fine-grained prosody transfer~(FPT)~\cite{klimkov2019finegrained,karlapati2020copycat,karlapati22copycat2}. FPT methods aim to extract prosody from a source speaker at a fine-grained level such as word, phoneme, or mel-spectrogram frame-level, and generate speech in a different target speaker's identity using prosody of the source speaker. Klimkov \emph{et~al.}~\cite{klimkov2019finegrained} showed that by extracting prosody relevant speech features like pitch, energy, and durations at the phoneme-level, we can transfer prosody at a fine-grained level from any speaker to a target speaker's identity. Karlapati \emph{et~al.}~\cite{karlapati2020copycat} used a conditional variational autoencoder to learn speaker-independent prosody representations at the frame-level to capture source prosody. In CopyCat2 (CC2), Karlapati \emph{et~al.}~\cite{karlapati22copycat2} showed that learning fine-grained speaker-independent prosody representations at the word-level helps reduce speaker leakage in FPT and can be used in the downstream task of TTS with contextually appropriate prosody. They did this by predicting the prosody representations learnt from speech using the contextual information available in the text. 
        
        In this work, we present eCat, with $4$ major contributions: i) to the best of our knowledge, eCat is the first end-to-end system capable of many-to-many FPT and multi-speaker TTS; ii) we show that training the acoustic model in an end-to-end fashion results in improved speaker similarity to the target speaker in FPT when compared to CC2; iii) we show that eCat provides naturalness improvements in multi-speaker TTS over CC2 on long-context text. These results are shown on English and Spanish internal datasets, consisting of data from en-US, en-GB, and es-US locales, and a total of 3 male and 4 female speakers; iv) we show that eCat is significantly preferred over VITS~\cite{vits-kim21f}, a model built on a different architecture, in multi-speaker TTS.
    
    \vspace{-1.5mm}
    \section{Proposed Method: eCat}
        \vspace{-1.5mm}
        eCat consists of $3$ components: 1) end-to-end acoustic model, 2) duration model, and 3) long-context flow-based prosody predictor called FlowCat. These 3 components are trained using a two-stage approach~\cite{Kathaka,CAMP,karlapati22copycat2}. In Stage~I, we learn word-level speaker-independent prosody representations from multi-speaker data, while in Stage~II, we learn to predict these representations using the contextual information available in text.
		\begin{figure*}
			\centering
            \includegraphics[width=0.97\linewidth]{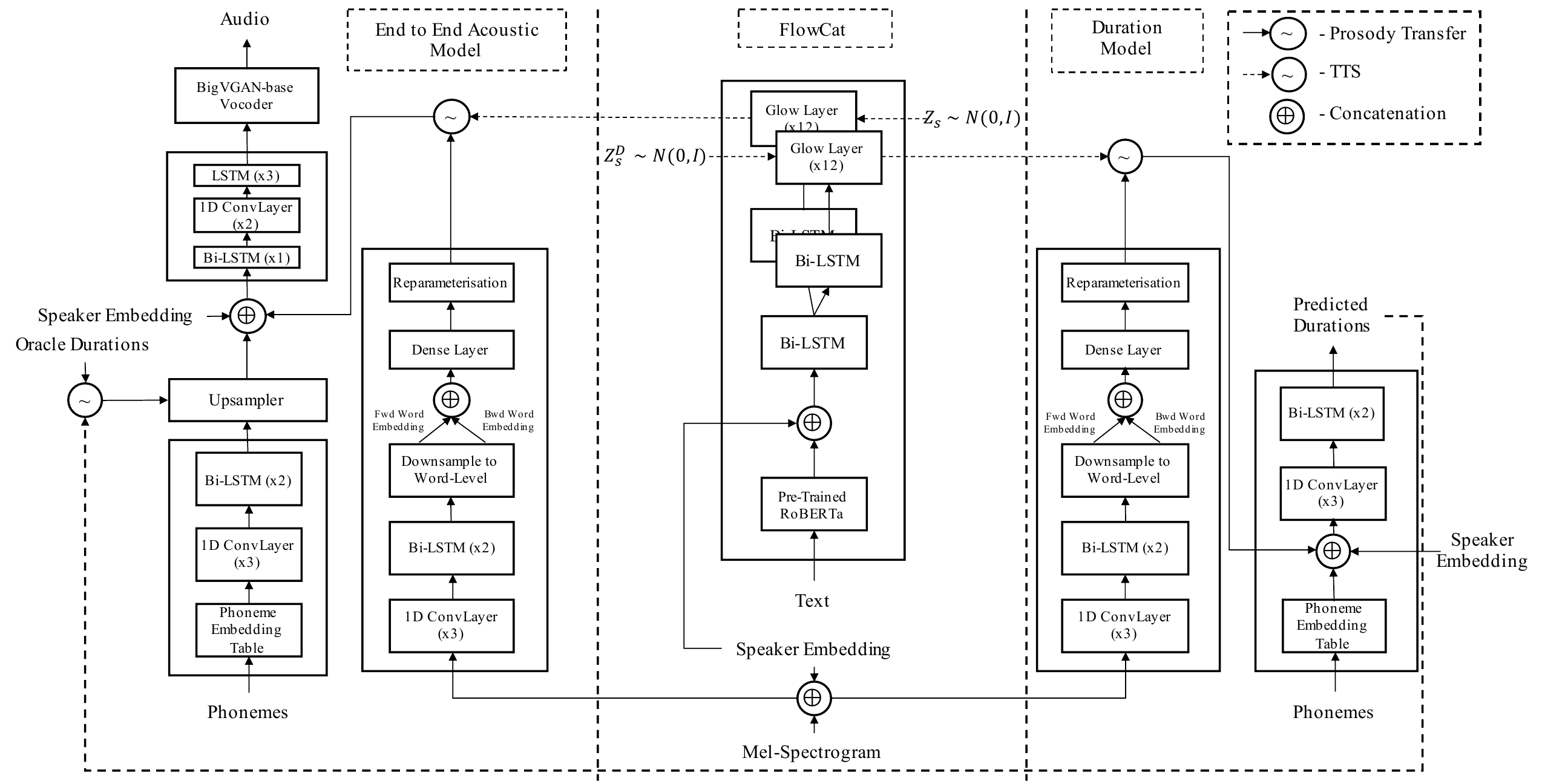}
            \vspace{-0.2cm}
            \caption{eCat architecture. Left: end-to-end acoustic model; middle: flow-based prosody predictor (FlowCat); right: duration model.}
            \label{fig:ecat}
            \vspace{-5mm}
		\end{figure*}
		
		\vspace{-5mm}
		\subsection{Stage~I: Learning Prosody Representations}
		    \vspace{-1mm}
		    \label{ssec:stage-I}
		    In eCat, as shown in Figure~\ref{fig:ecat}, the acoustic model consists of a phoneme encoder, a non-autoregressive (NAR) decoder, a conditional variational reference encoder, and the BigVGAN-base vocoder~\cite{lee2023bigvgan}. All components in the acoustic model are trained end-to-end. Similar to CC2, the phoneme encoder takes a vector of $P$ phonemes, $\vec{y}$, as input and provides phoneme encodings as output. These encodings are upsampled through replication according to the per-phoneme durations, $\vec{d}$, provided as input to the Upsampler. These upsampled encodings are passed to an NAR decoder along with speaker embeddings~\cite{karlapati2020copycat}, $\vec{c} \in \mathbb{R}^{E}$, where $E$ is the size of a speaker embedding. As the model is trained end-to-end, the decoder generates an intermediate representation, $B \in \mathbb{R}^{T_{mel} \times 80}$, instead of mel-spectrogram $X \in \mathbb{R}^{T_{mel} \times 80}$, where $T_{mel}$ is the number of mel-spectrogram frames. $B$ is then passed to the BigVGAN-base vocoder as input to generate a waveform, $\vec{x}$, of length $T$ samples. We aim to learn from speech a speaker-independent word-level representation of prosody, $Z = \lbrack \vec{z}_1, \vec{z}_2, \dots, \vec{z}_W \rbrack$, where $\vec{z}_i \in \mathbb{R}^U$ is a word-level prosody vector, $W$ is number of words in a sentence, and $U$ is the dimension of each prosody vector. As computing the true distribution of prosody is intractable, we use variational methods and the re-parameterisation trick to approximate the word-level distribution of prosody using a Conditional VAE, $q_\phi(Z \mid \vec{x}, \vec{y}, \vec{d}, \vec{c})$~\cite{sohn2015learning}. We assume that the vocoder parameters are also included in $\theta$ and train the end-to-end acoustic model to maximise the ELBO: $log\ p_\theta(\vec{x} \mid \vec{y}, \vec{d}, \vec{c}, Z) - KL(q_\phi(Z \mid \vec{x}, \vec{y}, \vec{d}, \vec{c}) \mid\mid p(Z))$. 
            
            In CC2, the authors had a similar ELBO where they used the $L2$ loss between generated mel-spectrograms and those obtained from speech waveforms as a proxy for negative log-likelihood. In eCat, we maximise the log-likelihood by using a modified GAN training scheme from BigVGAN~\cite{lee2023bigvgan}. We use the end-to-end acoustic model as the generator and represent the generated waveform as $\vec{x'}$. We use $5$ multi-periodicity ($D_{MPD}$) and $3$ multi-resolution ($D_{MRD}$) discriminators used in BigVGAN as the discriminators, $D \in \{D_{MPD}\} \cup \{D_{MRD}\}$. We use $\mathcal{A}$ to denote our dataset consisting of tuples of speech waveforms and corresponding text ($\vec{x}, t_x$). For each discriminator $D_k \in D$, the generator and discriminator losses are given as:
            \vspace{-1mm}
            \begin{equation}
                    L^{k}_{g} = \mathop{\mathbb{E}}_{Z \sim q_\phi} \Big[ (D_k(\vec{x'}) - 1)^2\Big],
                \vspace{-2mm}
            \end{equation}
            \begin{equation}
                    L^{k}_{d} = \mathop{\mathbb{E}}_{(\vec{x}, \vec{y}) \sim \mathcal{A},\ Z \sim q_\phi} \Big[(D_k(\vec{x}) - 1)^2 + (D_k(\vec{x'}))^2 \Big].
                \vspace{-2mm}
            \end{equation}
            We also use modified versions of the feature matching loss ($L_{f}$) and mel-spectrogram loss ($L_p$) from BigVGAN. We take the feature matching loss as the $L1$ loss between the outputs of each intermediate layer in a discriminator, $D^i_k \in D_k$, and use $\mid D_k \mid$ to refer to the number of layers in $D_k$. For $L_p$, we compute the $L1$ loss between the mel-spectrograms $X$ and $X'$ of the target and generated waveforms $\vec{x}$ and $\vec{x'}$:
            \vspace{-2mm}
            \begin{equation}
                \begin{split}
                    L^{k,i}_{f} &= \ \mid\mid D^i_k(\vec{x}) - D^i_k(\vec{x'}) \mid\mid_1 \\
                    L^{k}_{f} &= \mathop{\mathbb{E}}_{(\vec{x}, \vec{y}) \sim \mathcal{A},\ Z \sim q_\phi} \Bigg[\frac{1}{\mid D_k \mid} \sum_{i} L^{k,i}_{f} \Bigg], \\
                    L_{p} &= \mathop{\mathbb{E}}_{(\vec{x}, \vec{y}) \sim \mathcal{A},\ Z \sim q_\phi} \Big[\mid\mid X - X' \mid\mid_1 \Big].
                \end{split}
                \vspace{-2mm}
            \end{equation}
            We use $\eta$ to denote the parameters in all the discriminators. We know that $\theta$ and $\phi$ are the set of all acoustic model parameters. Thus, the acoustic model and discriminator losses are:
            \vspace{-1.5mm}
            \begin{equation}
                \begin{aligned}
                    \mathop{\mathrm{argmin}}_{\theta,\ \phi}\ \mathbb{L}_{AM} &= \sum_{k} \Big( L^{k}_{g} + \lambda_{f} L^{k}_{f} \Big) + \lambda_{p}\ L_{p} \\ & + \alpha\ \sum_{i=1}^{W} KL(q_\phi(\vec{z}_i \mid \vec{x}, \vec{c}) \mid \mid p(\vec{z}_i))\ , \\
                    \mathop{\mathrm{argmin}}_{\eta} \mathbb{L}_D &= \sum_{k} L^{k}_{d} \ .
                \end{aligned}
                \vspace{-1.5mm}
            \end{equation}
            
            As shown to the right in Figure~\ref{fig:ecat}, the duration model in eCat is unchanged from CC2. We learn word-level speaker-independent duration prosody representations, $Z^D \in \mathbb{R}^{W \times U^D}$, where $U^D$ is the size of each representation. To learn $Z^D$, like in the acoustic model we use a conditional variational reference encoder. We train the duration model, $r_\theta$ to maximise the evidence lower bound (ELBO): $log\ r_\theta(\vec{d} \mid \vec{y}, \vec{c}, Z^D) - KL(q_\psi(Z^D \mid \vec{x}, \vec{c}) \mid\mid r(Z^D))$, where $r(Z^D)=N(\vec{0}, I)$. 
            \vspace{-2mm}
            
        \subsection{Stage~II: FlowCat - Predicting Prosody Representations from Text}
            \label{ssec:stage-II}
            \vspace{-1mm}
            The acoustic and duration prosody distributions, $Z\sim q_\phi$ and $Z^D\sim q_\psi$ are learnt at the end of Stage~I. Both distributions are conditional upon $\vec{x}$, which is unavailable during inference. We learn to predict these latents using the contextual information available in text. As shown in the middle in Figure~\ref{fig:ecat}, we define a prosody predictor called \mbox{FlowCat}, $s_\nu(Z,\ Z^D \mid t_x, \vec{c})$, to predict $Z$ and $Z^D$ given text $t_x$ and speaker identity $\vec{c}$. We use normalising flows and condition them on the contextual information available in text. We use RoBERTa~\cite{liu19roberta} to get contextualised word embeddings from text. We combine these word embeddings with speaker embeddings to condition the flow layers. Normalizing flows are a class of generative models which provide exact log-likelihood estimates by using invertible functions and the change of variables trick~\cite{papamakarios21flows}. We use Glow-based flow layers as our flow functions~\cite{kingma18glow} which assumes conditional independence between word-level prosody representations.
            
            We note that the individual word-level prosody vectors $\vec{z}_i$ and $\vec{z}^D_i$ learnt in Stage~I are regularised to be from a simple distribution due to the prior being $N(0, I)$ in the KL divergence. However, we hypothesise that the sequence of word-level prosody latents $Z$ and $Z^D$ are complex distributions. Therefore, we conjecture that flows allow us to sample varied points from the complex prosody distributions, leading to expressive prosody in synthesised speech. When training, we use normalizing flows to transform a sequence of latent vectors from the target prosody distributions, $Z\sim q_\phi$ and $Z^D\sim q_\psi$ to a sequence of latent vectors from simple distributions $Z_{s} \sim \mathcal{N}(\vec{0}, I)$ and $Z^D_{s} \sim \mathcal{N}(\vec{0}, I)$. We do so through a composition of $K$ invertible flow functions $\vec{f} = f_1 \circ f_2 \circ \dots \circ f_K$. We define $H_i=f_i(H_{i-1};\ t_x, \vec{c})$ as the output from each flow function. Each $H_{i}=\lbrack \vec{h}_{1,\ i},\ \vec{h}_{2,\ i},\ \dots,\ \vec{h}_{W,\ i} \rbrack$ is a sequence of $W$ vectors. We define $H_0 = Z$, $H^D_0 = Z^D$, $H^D_K = Z^D_s$, $H_K = Z_{s}$, $s(Z_s) = \mathcal{N}(\vec{0}, I)$, and $s(Z^D_s) = \mathcal{N}(\vec{0}, I)$. We maximise the following log-likelihood of sampling points from the target distribution:
            \vspace{-3mm}
            \begin{equation}
                \begin{aligned}
                    &log\ s_\nu(Z,Z^D \mid t_x, \vec{c}) = \mathop{\mathop{\mathbb{E}}_{Z \sim q_\phi}}_{Z^D \sim q_\psi}\Bigg[ s(Z_{s}) + s(Z^D_s) + \\ &\sum_{i=1, w=1}^{K,W} log\ det\Bigg(\frac{\partial\vec{h}_{w,i}}{\partial\vec{h}_{w,i-1}}\Bigg) + log\ det\Bigg(\frac{\partial\vec{h}^D_{w,i}}{\partial\vec{h}^D_{w,i-1}}\Bigg) \Bigg].
                \end{aligned}
                \vspace{-3mm}
            \end{equation}
    \vspace{-3mm}
    \section{Experiments}
        \vspace{-1.5mm}
        \subsection{Data}
            \vspace{-1mm}
            We conducted experiments on 2 internal datasets ($\mathcal{A}_1$ \& $\mathcal{A}_2$). Both datasets consist of speakers reading excerpts from Wikipedia articles, news articles, etc. $\mathcal{A}_1$ is an English dataset and contains $80$ hours of speech from $4$ speakers: $1$ en-US female, $2$ en-US male, and $1$ en-GB female. $\mathcal{A}_2$ is an es-US dataset and contains a total of $40$ hours of speech from $1$ male and $2$ female speakers. All recordings were sampled at $\SI{24}{\kilo\hertz}$. We split each speaker's data in each dataset into train, validation, and test sets, in the ratio of $7:1:2$ without replacement.
            \vspace{-6mm}
        \subsection{Training \& Hyperparameters}
            \vspace{-1mm}
            The hyper parameters used to train our eCat models are shown in Table~\ref{tab:ecat_hparams}. In Stage~I, we trained the end-to-end acoustic model and the duration model without multi-sentence input. Training the end-to-end acoustic model to synthesise the whole sentence while generating the waveform at $\SI{24}{\kilo\hertz}$, is very slow. We make $2$ deviations to speed-up model training. First, we randomly sample chunks of length $M$ from the waveform to be synthesised during training. Second, we provide mel-spectrograms which are a compressed representation of speech to the reference encoder. We don't pass durations and phonemes to the reference encoder as this information is already captured in the mel-spectrograms. It creates word-level representations using frame-to-word alignments which are obtained via forced alignment. We trained the end-to-end acoustic model on $7$ V100 GPUs for $440$ epochs with a batch size of $84$.
            
            \begin{table}[t]
                \caption{Hyperparameters in eCat.}
                \vspace{-2mm}
                \label{tab:ecat_hparams}
                \centering
                \begin{tabular}{ r  r  r }
                    \toprule
                    \multicolumn{1}{c}{\textbf{Name}} & \multicolumn{1}{c}{\textbf{Symbol}} & \multicolumn{1}{c}{\textbf{Value}} \\
                    \midrule
                    Speaker Embeddings              &   $E$             &   $192$ dims       \\
                    Acoustic Word Prosody Latents   &   $U$             &   $4$ dims         \\
                    Duration Word Prosody Latents   &   $U^D$           &   $2$ dims         \\
                    Number of Glow Layers           &   $K$             &   $12$ layers      \\
                    Feature Matching Loss Weight    &   $\lambda_f$     &   $4$              \\
                    Mel-spectrogram Loss Weight     &   $\lambda_p$     &   $45$             \\
                    KL Divergence Weight            &   $\alpha$        &   $10^{-3}$        \\
                    Waveform Chunk Size             &   $M$             &   $19200$ samples  \\
                    \bottomrule
                \end{tabular}
                \vspace{-6.5mm}
            \end{table}
            
		    In Stage~II, we train FlowCat with multi-sentence input. We concatenate consecutive sentences to create sentences within a range of $72-95$ words, and provide this as input to \mbox{FlowCat}. We use the prosody embeddings obtained from Stage~I to act as the target for FlowCat. This provides our language model with extended context to obtain more contextually relevant word-embeddings. We used pre-trained RoBERTa-base models from Hugging Face~\cite{wolf20transformers} for $\mathcal{A}_1$~\cite{liu19roberta} and $\mathcal{A}_2$~\cite{rosa2022BERTIN}, and fine-tune them during training. We trained FlowCat on $4$ V100 GPUs for $106$ epochs with a batch size of $128$.
            \vspace{-2.5mm}
            
		\subsection{Inference}
		    \vspace{-1.5mm}
		    eCat has 2 inference modes: FPT and TTS. In FPT mode, we provide the source waveform and source speaker embeddings from which the prosody is extracted using the duration and acoustic reference encoders. Both the acoustic and duration models are then conditioned with input phonemes from the text and target speaker embedding to generate speech with the target speaker's identity and source prosody.

		    In TTS mode, we first run FlowCat on a window of text, including the target sentence to be synthesised and its surrounding context. FlowCat predicts the acoustic and duration prosody latents for all the words in the window. We only select latents for the target sentence, and use them in place of the outputs from the duration and acoustic reference encoders. The duration model then predicts per-phoneme target sentence durations. They are used by the end-to-end acoustic model along with the text-predicted acoustic word-level prosody latents, phonemes, and target speaker embeddings to synthesise speech.
            \vspace{-2mm}
            
        \subsection{Results}
            \vspace{-1.5mm}
            \subsubsection{Ablation Studies}
            \vspace{-1mm}
            \label{ssec:ablation}
            We conducted $2$ MUSHRA evaluations~\cite{itu20031534} on $\mathcal{A}_1$, to understand the separate contributions of end-to-end training with BigVGAN-base vocoder and FlowCat. Each MUSHRA was conducted with 24 listeners who were asked to rate, on a scale of 0 to 100, the naturalness of speech samples. Each sample had a concatenated duration of $20\sim30\text{secs}$. We used pair-wise two-sided Wilcoxon signed rank test with Bonferroni correction to measure the statistical significance of the results.
            
            First, we built a version of eCat, called E2E-CC2, where we used the Prosody Predictor from CC2 instead of FlowCat. We compared E2E-CC2 to CC2 and human recordings, making end-to-end training the only difference between the systems. As shown in Table~\ref{tab:ablation}, we found that E2E-CC2 is statistically significantly better than CC2. We hypothesise that this is due to end-to-end training solving the compounding errors problem, resulting in better segmental quality. Second, we compared E2E-CC2, eCat, and human recordings, to determine the impact of FlowCat. We also found that eCat is statistically significantly better when we use FlowCat instead of the prosody predictor from CC2. We conjecture that this is a result of more contextually appropriate and expressive speech due to the use of flows and better transitions between consecutive sentences due to multi-sentence context. Thus, both end-to-end training and FlowCat contribute to improving eCat.
            \begin{table}[pbt]
    		    \caption{{Mean MUSHRA scores for ablation study of end-to-end training \& FlowCat with $95\%$ CI on $\mathcal{A}_1$. (\textbf{Bold} indicates statistical significance)}}
    		    \vspace{-2mm}
                \centering
                \setlength{\tabcolsep}{3pt}
                \resizebox{\linewidth}{!}{%
                \begin{tabular}{lrrrr}
                \toprule
                Studied                  & \multicolumn{4}{c}{Mean MUSHRA scores} \\
                component                & CC2           & E2E-CC2          & eCat       & Human \\ 
                \midrule
                End-to-End               & $66.9\pm0.9$       & $\bf67.4\pm0.9$       & N/A        & $71.4\pm0.9$ \\
                FlowCat                  & N/A           & $71.3\pm1.5$          & $\bf73.9\pm1.4$ & $76.9\pm1.3$ \\
                \bottomrule  
                \end{tabular}
                }
                \label{tab:ablation}
                \vspace{-3mm}
            \end{table}
            
            \begin{table}[t]
    		    \caption{Mean MUSHRA scores in FPT for prosody similarity to source prosody and for speaker similarity to target speaker identity with $95\%$ CI on $\mathcal{A}_1$.}
    		    \vspace{-2mm}
                \centering
                \begin{tabular}{lrr}
                \toprule
                                &  \multicolumn{2}{c}{Mean MUSHRA scores} \\
                                & CC2 & eCat \\ 
                \midrule
                Prosody similarity    & $71.58\pm1.02$ & $71.70\pm1.03$ \\ 
                Speaker similarity    & $73.06\pm0.96$ & $\bf73.55\pm0.96$ \\ 
                \bottomrule  
                \end{tabular}
                \label{tab:fpt}
                \vspace{-6mm}
            \end{table}
            \vspace{-2.5mm}
            \subsubsection{Fine-grained Prosody Transfer}
                \vspace{-1mm}
                To evaluate FPT, we measured both the prosody similarity to the source prosody and the speaker similarity to the target speaker. We conducted $2$ MUSHRAs consisting of CC2 and eCat with a reference speech sample. When evaluating prosody similarity, the reference is a recording from the source speaker whose prosody is to be transferred. When evaluating target speaker similarity, we provide a recording in the target speaker's voice as reference. Each evaluation consisted of 100 listeners, each of whom was presented with 100 samples, 25 each from the speakers in $\mathcal{A}_1$. In evaluating prosody similarity, the listeners had to rate each sample on a scale of 0 to 100 based on how closely it followed the reference's prosody. As shown in Table~\ref{tab:fpt}, we found that eCat was on par with CC2~($\text{p-val}>0.05$). This shows that the prosody latent spaces learnt in both CC2 and eCat contain similar information for prosody transfer, implying that the improvements in prosody in TTS are owed to the improvement in prosody prediction, confirming FlowCat improvements from Section~\ref{ssec:ablation}. To evaluate speaker similarity, the listeners were asked to rate each sample on a scale of 0 to 100 based on how closely it resembles the voice identity of the reference speech. As shown in Table~\ref{tab:fpt}, eCat is statistically significantly better ($\text{p-val}<0.05$) than CC2 in terms of speaker similarity.
                
            \begin{table}[t]
                \caption{Mean MUSHRA scores for each speaker evaluated on TTS naturalness from $\mathcal{A}_1$ \& $\mathcal{A}_2$ dataset with $95\%$ CI.}
                \vspace{-2mm}
                \centering
                \begin{tabular}{lrrr}
                \toprule
                                 & \multicolumn{3}{c}{Mean MUSHRA scores} \\
                                 & CC2 & eCat  & Human \\ 
                \midrule
                \multicolumn{4}{c}{English ($\mathcal{A}_1$)} \\
                \midrule
                en-US male 1    & $68.5\pm1.4$ & $\bf73.1\pm1.4$ & $77.5\pm1.3$ \\ 
                en-US male 2    & $69.1\pm1.4$ & $\bf74.4\pm1.3$ & $78.9\pm1.2$ \\ 
                en-US female 1  & $70.3\pm1.4$ & $\bf73.1\pm1.4$ & $77.9\pm1.2$ \\ 
                en-GB female 1  & $74.6\pm1.3$ & $\bf76.7\pm1.2$ & $79.2\pm1.2$ \\ 
                \midrule
                \multicolumn{4}{c}{Spanish ($\mathcal{A}_2$)} \\
                \midrule
                es-US male 1    & $75.1\pm1.8$ & $\bf78.2\pm1.7$ & $81.8\pm1.6$ \\ 
                es-US female 1    & $71.0\pm1.8$ & $\bf74.8\pm1.6$ & $78.7\pm1.5$ \\ 
                es-US female 2  & $73.7\pm1.7$ & $\bf76.5\pm1.6$ & $80.1\pm1.5$ \\ 
                \bottomrule  
                \end{tabular}
                \label{tab:mushra}
                \vspace{-3mm}
            \end{table}
        
            \begin{table}[pbt]
    		    \caption{Percentage preferences in the preference test.}
    		    \vspace{-2mm}
                \centering
                \begin{tabular}{lrrr}
                \toprule
                                & VITS & No Preference & eCat \\ 
                \midrule
                en-US male 1    & 31.93\% & 32.53\% & 35.53\% \\
                en-US male 2    & 30.00\% & 31.80\% & \textbf{38.20\%} \\
                en-US female 1  & 29.67\% & 33.07\% & \textbf{37.27\%} \\
                en-GB female 1  & 30.40\% & 32.53\% & \textbf{37.07\%} \\
                \bottomrule  
                \end{tabular}
                \label{tab:compare_vits}
                \vspace{-6mm}
            \end{table}
            
            \vspace{-2.5mm}
            \subsubsection{TTS Naturalness}
                \vspace{-1mm}
                We trained separate models for $\mathcal{A}_1$ and $\mathcal{A}_2$ datasets. For each dataset, we conducted a MUSHRA evaluation between CC2, eCat, and human recordings. Both MUSHRAs were run on concatenated samples having a total duration of $20\sim30\text{secs}$. For $\mathcal{A}_1$, we randomly chose 25 such samples from the test set of each of the $4$ speakers leading to $100$ samples. We found that eCat statistically significantly reduces the gap between CC2 and human recordings ($\text{p-val}<0.05$) for each speaker, and by $46.9\%$ overall. For $\mathcal{A}_2$, we randomly chose $20$ concatenated samples from each speaker's test set for a total of $60$ samples. We found that eCat reduces the gap between CC2 and human recordings statistically significantly ($\text{p-val}<0.05$) for each speaker, and by $46.5\%$ overall. This is very close to the overall gap reduction observed on $\mathcal{A}_1$ as shown in Table~\ref{tab:mushra}. We hypothesise that this is because of improvements in both prosody and segmental quality. As shown in the ablation study in Section~\ref{ssec:ablation}, the improvement in prosody comes from FlowCat, leading to more expressive, coherent and contextually appropriate prosody in the synthesised speech, while the improvement in segmental quality is owed to the end-to-end training.
                
                We also conducted a preference test between eCat and VITS~\cite{vits-kim21f}. VITS is an end-to-end model with a different modelling approach that has demonstrated state-of-the-art results in naturalness of synthesised speech. We trained VITS using its public implementation~\footnote{\scriptsize\url{https://github.com/jaywalnut310/vits}} with BigVGAN-base vocoder instead of HiFi-GAN V1~\cite{kong2020hifigan} for a fairer comparison. The preference test was conducted on $\mathcal{A}_1$ dataset with $15$ concatenated samples for each of the $4$ speakers totalling to $60$ samples. As shown in Table~\ref{tab:compare_vits}, we found that eCat is statistically significantly preferred to VITS on $3$ out of $4$ speakers with a consistent improvement. On further analysis, we find that eCat has better expressiveness and consistent prosody across sentences when compared to VITS. This shows the importance of prosody prediction in improving the naturalness of synthesised speech. We also find that both systems have similar segmental quality as they both use the same vocoder and are trained end-to-end.
    
    \vspace{-2.5mm}
    \section{Conclusion}
        \vspace{-1mm}
        In this paper we introduced eCat, which to the best of our knowledge is the first end-to-end model capable of both fine-grained prosody transfer and multi-speaker TTS. We demonstrated that both end-to-end training and FlowCat bring statistically significant improvements through ablation studies. We showed that eCat reduces the gap in naturalness between CC2 and human speech in $3$ locales and $7$ voices by an average of $46.7\%$. We compared eCat to VITS, a state-of-the-art TTS model, and showed that eCat is statistically significantly preferred in $3$ out of the $4$ voices. We also showed that eCat improves target speaker similarity in prosody transfer settings in comparison to CC2.
        
    \vspace{-3mm}
    \section{Acknowledgements}
    \vspace{-1.5mm}
    We would like to thank Viacheslav Klimkov for his help with the implementation of end-to-end training mechanism of eCat.
    
    \newpage
    
    \bibliographystyle{IEEEtran}
    \bibliography{references}

\end{document}